\documentstyle[prd,aps,floats]{revtex}
\begin{document}
\draft
%
%
\input epsf
\renewcommand{\topfraction}{0.8}
\twocolumn[\hsize\textwidth\columnwidth\hsize\csname
@twocolumnfalse\endcsname
\preprint{CERN-TH/97-08, SU-ITP-97-04, astro-ph/9701173}
\title{Open Hybrid Inflation}
\author{Juan Garc\'{\i}a-Bellido}
\address{Theory Division, CERN, CH-1211 Geneva 23, Switzerland}
\author{Andrei Linde}
\address{Physics Department, Stanford University,
Stanford CA 94305-4060, USA}
\date{January 22, 1997}
\maketitle
\begin{abstract}
  We propose an open hybrid inflation scenario that produces an open
  universe with a `tilted' $n>1$ spectrum of metric perturbations. The
  model contains a symmetry breaking field that tunnels to its true
  vacuum, producing a single bubble inside which hybrid inflation
  drives the universe to almost flatness.  In order to obtain density
  perturbations with $n > 1$ we analyse a recently proposed new
  version of hybrid inflation scenario called tilted hybrid inflation.
  In this scenario, unlike in the previously known versions of hybrid
  inflation, a considerable tilt of the spectrum can be obtained
  without fine-tuning. The stage of inflation in this model is rather
  short, which allows us to obtain an inflationary universe with
  $\Omega < 1$ in a more natural way.  We study the separate
  contribution of scalar perturbations coming from the continuum
  subcurvature modes, the discrete supercurvature mode and the bubble
  wall mode to the angular power spectrum of temperature fluctuations
  in open inflation. We derive bounds on the parameters of the model
  so that the predicted spectrum is compatible with the observed
  anisotropy of the microwave background.
\end{abstract}
\pacs{PACS numbers: 98.80.Cq \hspace{8mm} Preprint CERN-TH/97-08,
SU-ITP-97-04, astro-ph/9701173}

\vskip2pc]

\section{Introduction}

Until recently, inflation was associated unequivocally with a flat
universe, due to its tendency to drive the spatial curvature so
effectively to zero. However, it is now understood that inflation
comprises a wider class of models, some of which may give rise to an
open universe at present~\cite{open,BGT,LM}. Such models generically
contain a field trapped in a false vacuum, which tunnels to its true
vacuum via the nucleation of a single bubble inside which a second
period of inflation drives the universe to almost flatness. These
models recently became rather popular because in an open universe it
is possible to reconcile a large value of the Hubble
constant~\cite{HST} with the large estimated age of globular
clusters~\cite{GC}. But even independently of this issue, open
inflation models have several interesting features that single them
out from other cosmological models.

For a long time it seemed impossible to make any physical sense of the
(mathematically consistent) open universe model because it presumed
that an infinite universe appeared from nowhere at a single moment of
time, being perfectly synchronized over an infinitely large length
scale. This leads to the horizon problem in its strongest form. Also,
whereas it seemed possible that a small closed universe could be
created by `tunneling from nothing' and then experience a period of
inflation, a similar event for an infinite open universe seemed
impossible. Indeed, one may argue that such processes should be
suppressed by $e^{-|S|}$, where $S$ is the action of the instanton
describing the universe creation \cite{LVIL}. Therefore one could
expect that quantum creation of an infinitely large open universe
should be forbidden because this would involve tunneling with infinite
action.  The models of Refs.~\cite{open,BGT,LM} show for the first time
that a consistent physical model of an infinite homogeneous universe
is possible, and that such a universe can appear as a result of
quantum tunneling. The bubble containing an open universe looks finite
from the point of view of an outside observer, and the probability of
its creation is finite. Meanwhile an observer inside the bubble would
see an infinite open universe.

Another distinct feature of such models is that the homogeneity
problem is solved not through the exponential expansion, as in the
usual inflationary models, but through the exponential suppression of
the probability of production of spherically asymmetric
bubbles~\cite{LM}. This way one solves the homogeneity problem
independently from the flatness problem. The origin of structure is
still related to amplified quantum fluctuations of the field that
drives inflation inside the bubble~\cite{sasaki,BGT}. However, in the
spectrum of temperature fluctuations there appears a new set of
discrete supercurvature modes~\cite{LW,GZ}, associated with the open
de Sitter vacuum~\cite{sasaki,YST} and the bubble wall fluctuations at
tunneling~\cite{LM,Hamazaki,bubble,Garriga}, which could be made small
in some of the models~\cite{super}.

There is some evidence that the observations made in a wide range of
scales, from horizon size to large clusters of galaxies, constrain
open inflation models (with small $\Omega_0\sim0.3$--$0.4$) to have a
`tilted' spectrum of density perturbations with the spectral index
$n>1$ ~\cite{WS}, and essentially no other contribution, either from
gravitational waves or supercurvature modes. The open inflation
models considered so far predict a tilted $n<1$ spectrum, see e.g.
Ref.~\cite{induced},  which may contradict observational data
for models with very small $\Omega_0$~\cite{WS}.

In this paper, we consider a model of open inflation in which the
second stage inside the bubble is driven by hybrid
inflation~\cite{hybrid}. Models of hybrid inflation are known to
provide a blue tilted spectrum of metric perturbations, with
negligible contribution of gravitational
waves~\cite{LL93,CLLSW,GBW,Lyth}. However, in most of the hybrid
inflation models developed so far the tilt is extremely small; in
order to achieve a considerable tilt of the spectrum in hybrid
inflation one needs to fine tune the parameters of the
model~\cite{GBW}. In what follows we will consider a new class of
hybrid inflation models where a significant tilt can be achieved in a
natural way, which we call {\it tilted hybrid inflation}, see
Ref.~\cite{GBL}.  We will make tilted hybrid inflation a part of an
open inflation scenario. To make sure that supercurvature and bubble
wall modes do not distort the CMB too much, we compute the angular
power spectrum of temperature anisotropies for both the continuum
modes and the supercurvature modes and find that there are models in
which all constraints are satisfied.

Apart from scalar metric perturbations, open inflation also produces a
primordial spectrum of gravitational waves, whose amplitude and scale
dependence in single-bubble open inflation models has only recently
been known~\cite{TS}. In order to compare with observations there
still remains to be computed the corresponding angular power spectrum
of CMB temperature fluctuations. We will consider the constraints on
the parameters of tilted hybrid models from such a tensor component of
the CMB anisotropies in a future publication.

\section{Tilted Hybrid Inflation}\label{DYN}

The simplest realization of hybrid inflation in the context of the
chaotic inflation scenario is provided by the potential~\cite{hybrid}
\begin{equation}\label{hybrid}
V(\phi,\psi) = {1\over4\lambda}\left(M^2-\lambda\psi^2 \right)^2
 + {1\over2}m^2\phi^2 + {1\over2}g^2\phi^2\psi^2 \, .
\end{equation}
The bare masses $m$ and $M$ of the scalar fields $\phi$ and $\psi$ are
`dressed' by their mutual interaction. At large values of the
fields, their effective masses squared are both positive and the
potential has the symmetry $\psi \leftrightarrow -\psi$. At small
values of the field $\phi$, the potential has a maximum at
$\phi=\psi=0$ and a global minimum at $\phi=0, \psi=\psi_0\equiv
M/\sqrt\lambda$, where the above symmetry is broken.

Equations of motion for the homogeneous fields at the stage of their
slow rolling during inflation are
\begin{eqnarray}\label{EQM}
3H\dot\phi &=& - (m^2 +g^2\psi^2) \phi\, , \\
3H\dot\psi &=& (M^2 - g^2\phi^2 - \lambda\psi^2) \psi \, ,
\end{eqnarray}
where the Hubble constant is given by $H^2 = {8\pi V/3M^2_{\rm P}}$.
Motion starts at large $\phi$, where the effective mass squared of the
$\psi$ field is positive and the field is sitting at the minimum of
the potential at $\psi=0$. As the field $\phi$ decreases, its quantum
fluctuations produce an almost scale invariant but slightly tilted
spectrum of density perturbations~\cite{hybrid,LL93,CLLSW,GBW,Lyth}.

During the slow-roll of the field $\phi$, the effective mass of the
triggering field is $m^2_\psi = g^2\phi^2 - M^2$. When the field
$\phi$ acquires the critical value $\phi_c \equiv M/g$, fluctuations
of the massless $\psi$ field trigger the symmetry-breaking phase
transition that ends inflation. If the bare mass $M$ of the $\sigma$
field is large compared with the rate of expansion $H$ of the
universe, the transition will be instantaneous and inflation will end
abruptly, as in the original hybrid inflation model~\cite{hybrid}. If
on the contrary the bare mass $M$ is of the order of $H$, then the
transition will be very slow and there is the possibility of having
yet a few more $e$-folds of inflation after the phase transition, see
Refs.~\cite{Guth,GBLW}.

When $\psi=0$ the inflaton potential becomes $V(\phi) = M^4/4\lambda +
m^2\phi^2/2$. Since the scalar field $\phi$ takes relatively small
values, for $m^2\ll g^2M^2/\lambda$ the energy density is dominated by
the vacuum energy,
\begin{equation}\label{H0}
H^2_0 = {2\pi M^4\over3\lambda M^2_{\rm P}}\,,
\end{equation}
to very good accuracy~\cite{GBW}. It is then possible to
integrate exactly the evolution equation of $\phi$,
\begin{eqnarray}\label{soln}
\phi(N) &=& \phi_c\,\exp(rN)\, , \\
r &=& {3\over2} - \sqrt{{9\over4}-{m^2\over H_0^2}} \,
\simeq {m^2\over3H_0^2}\,,
\end{eqnarray}
where $N=H_0(t_e-t)$ is the number of $e$-folds to the end of
inflation at $\phi=\phi_c$, and we have used the approximation
$m\lesssim H_0$.

An important feature of the perturbation spectrum of the hybrid
inflation model is its growth at large $k$; for example, at the end of
inflation in this theory one has \cite{hybrid,LL93}:
\begin{equation}\label{denspert}
\frac{\delta\rho}{\rho} (k) =  {2\sqrt{6\pi} g M^5\over 5\lambda
\sqrt{\lambda} m^2 M_{\rm P}^3} \cdot \left({k\over H_0}\right)^{m^2\over
3H_0^2}\ ,
\end{equation}
which corresponds to a spectral index
\begin{equation}\label{index}
n \simeq 1 + {2m^2\over 3H_0^2} = 1 + {\lambda m^2 M_{\rm P}^2
\over \pi M^4}\,.
\end{equation}
One should note that even though the spectral index $n$ in this model
is greater than 1, for typical values of the parameters, $n = 1$ with
great accuracy. For example, one may consider the numerical values of
parameters for the version of the hybrid inflation model discussed in
Ref.~\cite{hybrid}: $g^2 \sim \lambda \sim10^{-1}$, $m \sim 10^2$ GeV
(electroweak scale), $M \sim 10^{11}$ GeV. In this case $n-1= {\cal
  O}(10^{-4})$. Thus it is not very easy to use hybrid inflation in
order to obtain a blue spectrum without fine-tuning; in fact one could
even argue that hybrid inflation has a very stable prediction that the
spectrum must be almost exactly flat; see, however, Ref.~\cite{GBW}.

This conclusion could be somewhat premature. Indeed, one can follow
the lines of Ref.~\cite{LinMukh} and consider models where the field
$\phi$ during inflation acquires an effective mass squared
\begin{equation}\label{meff}
m^2_{\rm eff} = m^2 + \alpha H^2 \,,
\end{equation}
with $\alpha = {\cal O}(1)$. This is a very natural assumption, which
is true in a large class of supersymmetric models \cite{moduli}. It is
also true for models where the field $\phi$ has a non-minimal coupling
with gravity described by the additional term $-\xi R\phi^2/2$ in the
Lagrangian. During hybrid inflation $R = 12 H_0^2$, and the field
$\phi$ acquires an effective mass (\ref{meff}) with $\alpha = 12\xi $.
In such models one should consider those parts of the universe where
$\phi \ll M_{\rm P}/\sqrt\alpha$, because at much larger values of
$\phi$ the effective gravitational constant $G^{-1} = {M^2_{\rm P}/
16\pi-\xi\phi^2/2}$ becomes singular.  For $\phi \ll M_{\rm P}/
\sqrt\alpha$ the Hubble constant can be approximated by Eq.~(\ref{H0}).
In such a case one has
\begin{equation}\label{newden}
\frac{\delta\rho}{\rho} (k) =  {9\sqrt 2 H\over 5\pi \alpha\phi }  =
{3\sqrt{6}\, g M \over 5 \sqrt{\lambda \pi}\, \alpha M_{\rm P}  }
\left({k\over H_0}\right)^{\alpha\over3} \,.
\end{equation}
Here $\phi$ is the value of the scalar field at the moment when the
perturbations are produced which at the end of inflation have momentum
$k$.  These perturbations have spectral index $n - 1 \simeq 2\alpha/3
> 0$ for $\alpha \lesssim 1$. We call such models {\it tilted hybrid
  inflation}~\cite{GBL}: they provide a spectrum of density
perturbations with a positive tilt, which may be large and may not
require any fine tuning of parameters.

We will concentrate here, for definiteness, on a particular version of
the tilted hybrid model proposed in Ref.~\cite{GBL}. This model has a
potential for the field $\phi$ (while $\psi = 0$, i.e. during the
stage of inflation in our scenario), which is given by
\begin{equation}\label{POT}
V(\phi) = V_0 \,\exp\Big({4\pi\alpha\phi^2\over3M_{\rm P}^2}\Big)\,,
\end{equation}
where $V_0=M^4/4\lambda$ is the vacuum energy, and we are assuming
$4\pi\alpha\phi^2/3 \ll M_{\rm P}^2$. The effective mass of $\phi$
becomes
\begin{equation}\label{MAS}
m^2 = V''(\phi) = \alpha H^2\,\Big(1 +
{8\pi\alpha\phi^2\over3M_{\rm P}^2}\Big) \simeq \alpha H^2\,.
\end{equation}
The field evolves according to Eq.~(\ref{soln}) with $r\simeq
\alpha/3$, and $\dot\phi^2 + m^2\phi^2 \ll V_0$ is satisfied during
inflation. In this model, the condition for inflation to occur, $-\dot
H < H^2$, is
\begin{equation}\label{infl}
\phi < \phi_{\rm inf} \equiv {3\over2\sqrt\pi\,\alpha}\,M_{\rm P}\,.
\end{equation}
As a consequence, the number of $e$-folds has a maximum value
\begin{equation}\label{Nmax}
N_{\rm max} \equiv {3\over\alpha}\,\ln\Big({\phi_{\rm inf}\over
\phi_c}\Big) = {3\over\alpha}\,\ln\Big(
{3g\,M_{\rm P}\over2\sqrt\pi\,\alpha M}\Big)\,.
\end{equation}

\begin{figure}[t]
\centering
\leavevmode\epsfysize=5.5cm \epsfbox{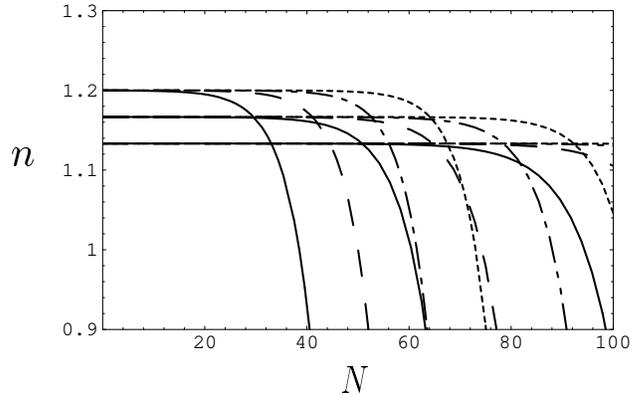}\\
\caption[fig1]{\label{fig1} The tilt of the spectrum (\ref{tilt})
  as a function of the number of $e$-folds to the end of inflation,
  for $N_{\rm hor}=55$ and parameters $\alpha = 0.3, 0.25, 0.2$,
  from top to bottom. The black, dashed, dotted-dashed and dotted
  lines for each $\alpha$ correspond to the ratio $g^4/\lambda = 0.1,
  1.0, 10, 100$, respectively.  Note that the tilt rises very quickly
  to its asymptotic value.}
\end{figure}

In this model, the spectrum of curvature perturbations for a flat
universe is expressed by~\cite{GBL}
\begin{equation}\label{PR}
{\cal P}_{\cal R}(k) = {9H^2(\phi)\over4\pi^2\alpha^2\phi^2}\,.
\end{equation}
In a flat universe, the COBE normalization imposes a
constraint on the amplitude (\ref{PR}), see Ref.~\cite{GBL},
\begin{equation}\label{scon}
{3g^2 M^2 \,e^{-2\alpha N/3}\over2\lambda\,\alpha^2\,M_{\rm P}^2}
\simeq 10^{-8}\,.
\end{equation}
where $N$ corresponds to the scale of the horizon today,
\begin{equation}
N_{\rm hor} \simeq 55 + {2\over3} \ln\Big({V^{1/4}\over10^{16}{\rm
GeV}}\Big)+ {1\over3} \ln\Big({T_{\rm rh}\over10^{9}{\rm GeV}}\Big)\,.
\end{equation}
As we will see in Section IV, the constraint in an open model is not
very different. For $g^2\sim\lambda\sim0.1$ and $\alpha\simeq 0.25$ we
have $M \simeq 2\times 10^{-3} M_{\rm P} \simeq 2 \times 10^{16}\,{\rm
  GeV}$, i.e. the GUT scale.

The spectrum (\ref{PR}) has a tilt given by
\begin{equation}\label{tilt}
n-1 = {d\ln{\cal P}_{\cal R}\over d\ln k} =
{2\alpha\over3}\Big(1 -
{4\pi\alpha\phi^2\over3M_{\rm P}^2}\Big)\,,
\end{equation}
which vanishes at
\begin{equation}\label{min}
\phi = \phi_{\rm min} \equiv \sqrt{3\over4\pi\alpha}\,M_{\rm P}\,,
\end{equation}
where the spectrum has a minimum. Later on, the tilt will acquire the
asymptotic value $n_{\rm as} = 1+2\alpha/3$. This is reflected in
Fig.~1, where it is seen how the tilt grows as a function of the
number of $e$-folds to the end of inflation, for various values of
$\alpha$. As we increase the ratio $g^4/\lambda$, the asymptotic tilt
is more quickly approached.

Using the constraint (\ref{scon}) it is possible to find an expression
for $N_{\rm max}$, as a function of $N_{\rm hor}$,
\begin{equation}
N_{\rm max} \simeq {12\over\alpha} \ln 10 - N_{\rm hor} +
{3\over2\alpha} \ln\Big({27\over8\pi\alpha^4}{g^4\over\lambda}\Big)\,.
\end{equation}

\begin{figure}[t]
\centering
\leavevmode\epsfysize=6.1cm \epsfbox{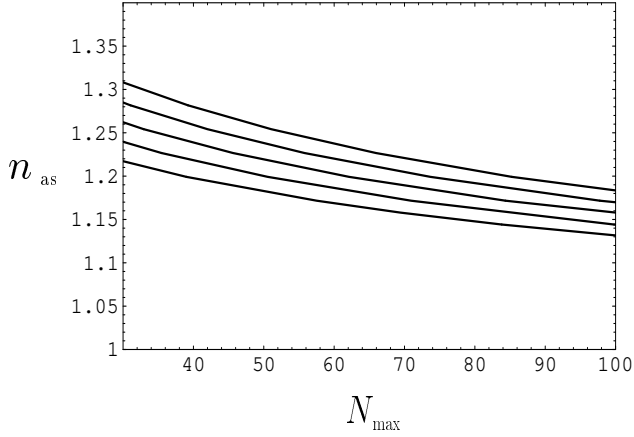}\\
\caption[fig2]{\label{fig2} The asymptotic tilt of the spectrum as a
  function of the maximum number of $e$-folds, $N_{\rm max}$, for
  $N_{\rm hor}=55$. The five lines correspond to the ratio
  $g^4/\lambda = 100, 10, 1, 0.1, 0.01$, from top to bottom.}
\end{figure}

In Fig.~2 we have plotted the asymptotic value of the tilt $n_{\rm as}
= 1 + 2\alpha/3$ as a function of $N_{\rm max}$, for various values of
$g^4/\lambda$. As we increase this ratio, larger asymptotic tilts
can be obtained for smaller values of $N_{\rm max}$. Thus strongly
tilted hybrid models generically predict a small number of $e$-folds
of inflation. This is another feature of the model that could be
useful for open inflation. For example, depending on the parameters of
the theory and the efficiency of reheating, the minimum of the
spectrum of curvature perturbations (\ref{min}) for $\alpha \sim 0.25$
could lie on scales comparable to the scale of our present horizon. If
the minimum corresponds to the horizon scale, the spectrum on this
scale will be practically flat.  For $M \simeq 2\times 10^{-3} M_{\rm
  P} \simeq 2 \times 10^{16}\,{\rm GeV}$, $\alpha=0.25$ and $g^2\simeq
\lambda \simeq 0.1$, we find $N_{\rm max} \sim 60$, which is of the
order of the number of $e$-folds corresponding to the scale of the
horizon.  In this case, the scalar spectrum is nearly scale invariant,
see Eq.~(\ref{tilt}), on those scales, but soon rises to an asymptotic
value $n_{\rm as}\simeq 1.17$ on smaller scales.
This choice of parameters~\cite{GBL} is not unique; e.g. one can have a
larger $\alpha=0.3$, for $g=1$ and $\lambda=0.01$, which gives
$M = 6\times 10^{15}$ GeV and $n=1.2$ on all scales within the horizon,
as you can see from Fig.~1. In this case, $N_{\rm max} = 85$, which
is much larger than $N_{\rm hor} = 55$.

\section{Open Hybrid Inflation}

The new ingredient here is a period of false vacuum inflation prior to
the stage of hybrid inflation. This could be accounted for by a
symmetry breaking field $\sigma$ which is trapped in a   false
vacuum and eventually tunnels through to the true vacuum via the
nucleation of a bubble.  One of the possible models describing this process
has the following effective potential  ~\cite{LM}:
\begin{equation}\label{tunnel}
V(\sigma,\phi) = {M'^2 \sigma^2\over2} - \sqrt{\lambda'} M'\sigma^3 +
{ \lambda'\sigma^4\over4} + {h^2\phi^2\sigma^2\over2}  + \tilde V\,,
\end{equation}
where we have included a coupling between $\sigma$ and $\phi$ to
ensure the synchronization of the two fields after tunneling, in order
to prevent large temperature fluctuations in the CMB~\cite{LM}. The constant
$\tilde V \sim M'^4/\lambda'$ is added to make  the effective potential
vanish in its global minimum.

The combined effective potential for all fields in our model looks as follows:
\begin{eqnarray}\label{tunnel1}
 V(\sigma,\phi,\psi) &=&   -{M^2 \psi^2\over 2}+{\lambda\psi^4 \over 4 }
 + {M^4\over 4\lambda} \,\exp\Big({4\pi\alpha\phi^2\over3M_{\rm P}^2}\Big) +
\nonumber \\
&+ & {M'^2 \sigma^2\over2} - \sqrt{\lambda'} M'\sigma^3 +
{ \lambda'\sigma^4\over4}\nonumber \\
 &+& { \phi^2\over2}(g^2\psi^2 + h^2\sigma^2) + \tilde V\,.
\end{eqnarray}
Even though this model may look complicated, it is basically very
simple: $ V(\sigma,\phi,\psi) $ is the sum of the effective potential
for the tilted hybrid inflation and the potential (\ref{tunnel}) for
open inflation. The global minimum of this potential is at $\phi=0$,
$\psi = M/\sqrt \lambda$, $\sigma = {1\over2}(3+ \sqrt 5)\,
M'/\sqrt{\lambda'}$; the effective potential vanishes there. We
suppose that $ {M\over \sqrt \lambda}, {M'\over \sqrt{\lambda' }},
{M\over g}, {M'\over h} \ll M_{\rm P}$.

Now let us study the combined evolution of all three fields. Suppose
that initially the scalar field $\phi$ is very large, $\phi \gg M/g,
M'/h$. At that time the minimum of the effective potential is at $\psi
= \sigma = 0$, and the fields roll there.  The minimum at $\sigma\neq
0$ appears and becomes deeper than the minimum at $\sigma=0$ only for
$\phi<\phi_*\equiv M'/h$, and the phase transition to $\psi \not = 0$
occurs at $\phi < \phi_c \equiv M/g$.  Let us consider the model with
$M'/h > M/g$, i.e.  $\phi_* > \phi_c $.  In this case the first phase
transition occurs when the field $\phi$ rolls down to $\phi_*$, and
the bubbles of the field $\sigma$ containing open universes inside
each of them are formed.  Inside each bubble, the scalar field $\phi$
continues rolling down, driving the second stage of inflation which
eventually ends when the field $\phi$ reaches $\phi_c$. If the
tunneling event is very improbable, then other bubbles will not collide
with ours in our past light cone and we could be living inside such a
bubble.

The second stage of inflation inside each of the bubbles, which occurs
when the field $\phi$ rolls down from $\phi_c$ to $\phi_*$, can be
approximately described by the tilted hybrid inflation model described
in the previous section.
The number of $e$-folds of hybrid inflation
inside the bubble is then given by
\begin{equation}\label{efolds}
N \approx {3\over\alpha}\,\ln{\phi_*\over\phi_c} =
{3\over\alpha}\,\ln{g\,M'\over h\,M}\,.
\end{equation}
For an open universe we require $N=55\pm5$.

However, one should take into account also an additional contribution
of the term $h^2\phi^2\sigma^2/2 \sim h^2 \phi^2 M'^2/2\lambda'$ to the
effective mass of the scalar field $\phi$, as well as the contribution
of the false vacuum energy density of the field $\sigma$ to the Hubble
constant.
Fortunately, under the following conditions these corrections are
small and can be neglected:
\begin{eqnarray}\label{COND}
{h^2 M'^2\over \lambda'} &\ll&
{2\pi\alpha M^4\over3\lambda M_{\rm P}^2} \,,
\nonumber\\
{  M'^4\over \lambda'} &\ll& { M^4 \over  \lambda } \,.
\end{eqnarray}
Using Eqs.~(\ref{scon}) and (\ref{efolds}), the first condition can
be written as
\begin{equation}\label{con}
{h^4\over\lambda'} \ll {8\pi\alpha^3\over9}\,10^{-8}\,.
\end{equation}
All conditions will be satisfied if, for instance, one takes the values
$\alpha=0.25$, $g^2 = \lambda=0.1$ and $M = 2\times 10^{16}$ GeV, as in the
previous
Section, and  $\lambda' {\
\lower-1.2pt\vbox{\hbox{\rlap{$>$}\lower5pt\vbox{\hbox{$\sim$}}}}\ } 0.1$, $h
\sim 10^{-3}$, and $M'\sim
10^{16}$ GeV. In such a case all the results of the previous section
remain valid for our model. Thus we have found a working model of open
universe that produces metric perturbations with a blue spectrum, as
described in Sect.~\ref{DYN}.

\section{Temperature Anisotropies}

Until recently, observations of the CMB anisotropies provided just a
few constraints on the parameters of inflationary models, mainly from
the amplitude and the tilt of both scalar and tensor perturbations'
spectrum. Nowadays, with several experiments looking at different
angular scales, we have much more information about both spectra as
well as other cosmological parameters such as $\Omega_0$, $H_0$,
$\Omega_{\rm B}$, etc. In the near future, most of these parameters
will be known with better than a few \% accuracy, thanks to the new
generation of microwave anisotropy satellites, {\it MAP}~\cite{MAP}
and {\it COBRAS/SAMBA}~\cite{COBRAS}.

Quantum fluctuations of the inflaton field $\phi$ during inflation
produce long-wavelength scalar curvature perturbations and tensor
(gravitational wave) perturbations, which leave their imprint in the
CMB anisotropies. We will only consider here the scalar component and
leave the tensor for a future publication. We will use ${\cal R}$ to
denote the scalar metric perturbation on comoving hypersurfaces. Open
inflation generates three different types of scalar modes: those that
cross outside during the second stage of inflation and constitute a
continuum of subcurvature modes~\cite{sasaki,BGT}; a discrete
supercurvature mode~\cite{LW}, associated with the open de Sitter
vacuum~\cite{YST}, and a mode associated with the bubble wall
fluctuations at tunneling~\cite{bubble,Garriga,Cohn}, all of which
induce temperature anisotropies in the microwave background.

In order to compare with observations we have to compute the effect
that such a scalar metric perturbation has on the temperature
anisotropies of the CMB (expanded in spherical harmonics),
\begin{equation}\label{temp}
{\delta T\over T}(\theta,\phi) = \sum_{lm} a_{lm} \,
Y_{lm}(\theta,\phi)\,.
\end{equation}
In the context of an open universe, the main effect on large scales
comes from the integrated Sachs-Wolfe effect~\cite{SW67}. The angular
power spectrum $C_l\equiv\langle|a_{lm}|^2\rangle$ has contributions
coming from the continuum of scalar subcurvature modes, the
supercurvature mode and the bubble wall mode,
\begin{equation}\label{CL}
C_l = C_l^{(C)} + C_l^{(S)} + C_l^{(W)} \,.
\end{equation}
The contribution of each mode to the $C_l$'s is measured by a
window function $W_{ql}$, given by~\cite{SW67}
\begin{equation}\label{window}
5\,W_{ql} = \Pi_{ql}(\eta_0) + 6\int_0^{\eta_0} dr\,F'(\eta_0-r)\,
\Pi_{ql}(r)\,,
\end{equation}
for the subcurvature modes, where
\begin{equation}\label{F}
F(\eta)=5\,{\sinh^2\eta-3\eta\,\sinh\eta+4(\cosh\eta-1)\over
(\cosh\eta-1)^3}\,,
\end{equation}
and $\,\eta_0=\cosh^{-1}(2/\Omega_0-1)$ is the distance to the last
scattering surface. Here $\Pi_{ql}(r)$ are the radial eigenfunctions
of the Laplacian, with eigenvalue $-k^2=-(1+q^2)$, see
Ref.~\cite{Har}. The curvature scale corresponds to $k=1$. For
subcurvature modes, $q^2>0$, while for supercurvature modes,
$q^2<0$.  There is an analogous expression to (\ref{window}) for the
supercurvature modes' window functions $\bar W_{|q|l}$ in terms of
the eigenfunctions $\bar\Pi_{|q|l}(r)$.  The corresponding expressions
can be found in the Appendix of Ref.~\cite{induced}, where a similar
analysis was performed for the induced gravity open inflation model.

\subsection*{Subcurvature modes}

Those fluctuations that cross the Hubble scale during the second stage
of inflation give rise to a continuum spectrum of subcurvature scalar
metric perturbations~\cite{open,YST} which can be computed in the case
of hybrid inflation
\begin{eqnarray}\label{spectrum}
{\cal P}_{\cal R} &=&
A^2_C\,\coth(\pi q) \\[1mm]
A^2_C & = & {3g^2M^2\,e^{-2\alpha N/3}
\over2\pi\lambda\,\alpha^2 M_{\rm P}^2}\,.
\end{eqnarray}
The extra $\,\coth\,$ factor is due to the spatial curvature
immediately after tunneling~\cite{sasaki,YST}. As inflation progresses
inside the bubble, and the corresponding scales acquire values
$q\gg1$, the spectrum recovers the usual power law behavior of flat
inflation. The spectral tilt can then be evaluated as in
Eq.~(\ref{tilt}), $n=1+2\alpha/3$. Note that the tilt is always
greater than 1 in this model.

\begin{figure}[t]
\centering
\leavevmode\epsfysize=5.63cm \epsfbox{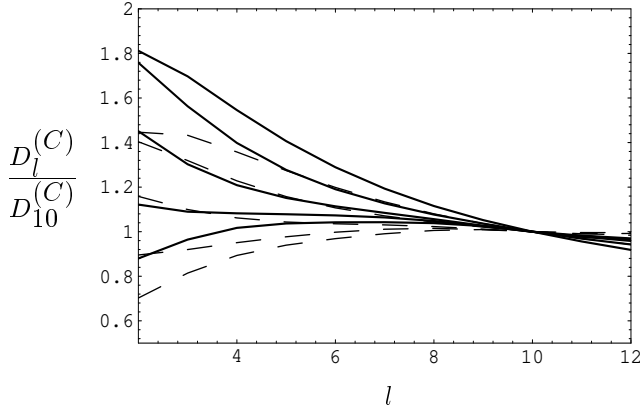}\\
\caption[fig3]{\label{fig3} The first 12 multipoles of the angular power
  spectrum associated with the continuum modes, $D_l^{(C)}$,
  normalized to the tenth multipole, for $\Omega_0 = 0.2$ to $0.6$
  from top to bottom at low $l$. The thick lines correspond to a
  scale invariant $n=1$ spectrum, while the thin dashed lines
  correspond to a tilted $n=1.15$ spectrum. }
\end{figure}

The angular power spectrum for the continuum modes can be
written as~\cite{sasaki}
\begin{equation}\label{powerC}
C_l^{(C)} = 2\pi^2\,\int_0^\infty
{dq\over q(1+q^2)}\,{\cal P}_{\cal R}(q)\,W_{ql}^2\,,
\end{equation}
where ${\cal P}_{\cal R}(q)$ is the primordial curvature
perturbations' spectrum (\ref{spectrum}) and the window function
$W_{ql}$ can be computed from Eq.~(\ref{window}) with eigenfunctions
given in Ref.~\cite{Har,induced}.

We can compute the large angle power spectrum for different values of
$\Omega_0$ in the low density range, $0.2\leq\Omega_0\leq0.6$, for a
scale invariant $n=1$ spectrum, as well as for a tilted, $n=1.15$,
spectrum. In Fig.~3 we show the first twelve multipoles, adopting the
notation $D_l = l(l+1)\,C_l$. Note that the quadrupoles of the two
spectra actually differ by a factor as large as $1.25$, for any value
of $\Omega_0$. Clearly, a tilted $n>1$ density perturbation spectrum
produces a flatter power spectrum in an open universe. We have
computed the large scale power spectrum including only the Sachs-Wolfe
effect, as appropriate to the lowest multipoles, while ignoring the
effect of oscillations in the photon-baryon fluid, which is known to
induce and effective extra tilt of about 0.15, as it rises to the
first acoustic peak at~\cite{peak}
\begin{equation}\label{lpeak}
l_{\rm peak} \simeq 220 \,\Omega_0^{-1/2}\,.
\end{equation}
A future precise determination of $l_{\rm peak}$ will allow us to
conclude whether the universe is indeed open~\cite{future}.

In order to compare our predictions with observations we have to know
the amplitude and shape of the power spectrum of temperature
fluctuations in the microwave background. The normalization to
COBE~\cite{COBE} for tilted open models has recently been given by
Bunn and White~\cite{BW}, under the assumption that only the continuum
modes are important. They specify a quantity $\delta_{\rm H}$, which
measures the normalization of the present matter power spectrum in an
open universe. The preferred value depends on $n$ and
$\Omega_0$,~\cite{BW}
\begin{eqnarray}\label{DH}
\delta_{\rm H}(n,\Omega_0) &=& 1.95\times10^{-5}\,
\Omega_0^{-0.35-0.19\ln\Omega_0-0.17(n-1)}\times \nonumber\\[1mm]
&&\hskip5mm \exp[-(n-1)-0.14(n-1)^2]\,.
\end{eqnarray}
We will consider values of the tilt away from 1, as well as low
density universes, so both dependencies in principle need to be taken
into account. However, the dependencies are not that strong for the
values we are considering, so it is enough to take $\delta_{\rm
  H}\simeq 2\times 10^{-5}$, in order to determine the model's
parameters. In the future, a precise knowledge of both $\Omega_0$
and $n$ will allow a better determination of the parameters. In an
open universe $\delta_{{\rm H}}$ is related to ${\cal P}_{\cal R}$
as~\cite{BW}
\begin{equation}
\delta_{\rm H} = {2\over5} \,{\cal P}_{\cal R}^{1/2}
        \,{g(\Omega_0)\over\Omega_0} \,,
\end{equation}
where $g(\Omega_0)$ is a function measuring the suppression in the
growth of perturbations relative to a critical-density universe,
\begin{equation}
{g(\Omega)\over\Omega} = {5\over2}\Big[{1\over70} +
{209\Omega\over140} - {\Omega^2\over140} + \Omega^{4/7}\Big]^{-1}\,,
\end{equation}
which tends to 1 as $\Omega_0\to1$, and ${\cal P}_{\cal R}$ is
evaluated at around the 10th multipole, where cosmic variance is small
and $\coth\pi q \simeq1$. The $\Omega_0$ dependence can give a factor
of up to 1.5 in the region of interest, but we can ignore it as we do
not require such accuracy.  Reproducing the amplitude of temperature
anisotropies gives the most important constraint on the parameters of
the model,
\begin{equation}\label{scalar}
{3gM\,e^{-\alpha N/3} \over2\alpha\sqrt{\lambda}\,M_{\rm P}}\,
\simeq 10^{-4}\,.
\end{equation}
For $N\simeq 55$ on the scale of the horizon, and very natural
values of the couplings, $g^2 \sim \lambda \sim 0.1$, we have
\begin{equation}\label{bound}
M \simeq 2\times 10^{-3} M_{\rm P} \simeq 2 \times 10^{16}\,{\rm GeV}\,,
\end{equation}
the GUT scale. If the scalar field $\phi$ has a bare mass $m$, it
should be smaller than
\begin{equation}\label{mass}
\sqrt\alpha H_0  \simeq 10^{-5} M_{\rm P}   \,.
\end{equation}
As long as $m$ is smaller than $\sqrt \alpha H_0$, nothing depends on
this parameter, so we do not need to fine-tune $m$ in order to obtain
adiabatic scalar perturbations with a blue spectrum. Since $H_0\ll
M$, hybrid inflation will end with a sudden transition to the symmetry
breaking phase of the triggering field, without a second stage of
inflation like the one discussed in Refs.~\cite{Guth,GBLW}.

\subsection*{Supercurvature mode}

Apart from the continuum of subcurvature modes, in open inflation we
also have a contribution to the microwave background anisotropies
coming from a discrete supercurvature mode,
\begin{equation}\label{super}
k^2 = 1 - \Big[\Big({9\over4}-{m_F^2\over H_F^2}\Big)^{1/2} -
{1\over2}\Big]^2\,,
\end{equation}
which appears in the spectrum of the inflaton field in open de Sitter
when $m_F^2<2 H_F^2$ in the false vacuum~\cite{sasaki}. The metric
perturbation for this mode is given by~\cite{super}
\begin{equation}\label{RS}
\langle{\cal R}^2_S\rangle \simeq A^2_C  \, {H_F^2\over H_T^2} \,.
\end{equation}

\begin{figure}[t]
\centering
\leavevmode\epsfysize=5.78cm \epsfbox{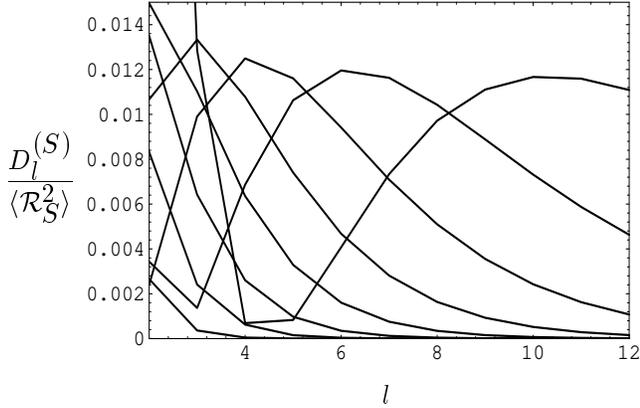}\\
\caption[fig4]{\label{fig4} The first 12 multipoles of the angular power
  spectrum associated with the supercurvature modes, $D_l^{(S)}$,
  normalized to the curvature perturbation amplitude, $\langle{\cal
  R}^2_S\rangle$, for $\Omega_0 = 0.2$ to $0.9$ from top right to
  bottom left. The quadrupole does not necessarily dominate
  the power spectrum for small $\Omega_0$.}
\end{figure}

The angular power spectrum of temperature aniso\-tropies induced
by this supercurvature mode can be written as
\begin{equation}\label{powerS}
D_l^{(S)} = l(l+1) C_l^{(S)} =
4\pi\,\langle{\cal R}^2_S\rangle\,\bar W_{1l}^2\,,
\end{equation}
where $\bar W_{1l}$ is the window function for this $|q|=1$ mode, see
Ref.~\cite{induced}, where the correctly normalized eigenfunctions
were given.

We have computed the angular power spectrum $D_l^{(S)}$ for different
values of $\Omega_0$ in the range $0.2\leq\Omega_0\leq0.9$, as shown
in Fig.~4 for the first twelve multipoles. We have found
$D_l^{(S)}/\langle{\cal R}^2_S\rangle$ to be two
orders of magnitude smaller than the continuum spectrum, see Fig.~6.
Since we want the supercurvature mode contribution to the power
spectrum to be smaller than that of the continuum modes, we need
\begin{equation}\label{condS}
\langle{\cal R}^2_S\rangle < 100\, A^2_C \,.
\end{equation}
This imposes only a very mild constraint on the parameters of the
model,
\begin{equation}\label{constrS}
{H_F^2\over H_T^2} <  100\,,
\end{equation}
which is easy to satisfy for reasonable values of the parameters.
Since we are considering that the tunneling occurs at $\phi \simeq
\phi_*$, where the two minima of the tunneling potential are
degenerate, we have $H_F\simeq H_T$ and therefore the bound is easily
satisfied.

\subsection*{Bubble wall mode}

Apart from the continuum of sub\-curvature modes and the dis\-crete
super\-curvature mode, we expect also a contribution from the bubble
wall fluctuations~\cite{bubble,Garriga,Cohn}. These fluctuations
contribute as a transverse traceless curvature perturbation $k^2=-3$
mode, which nevertheless behaves as a homogeneous random field, see
Refs.~\cite{Hamazaki,modes}.

The curvature perturbation amplitude for this bubble wall mode can be
computed from~\cite{bubble,YST}
\begin{equation}\label{wall}
\langle{\cal R}^2_W\rangle = {3\kappa^2\,H_T^2\over16\pi^2\,
\alpha^2\beta}\,\Big[\alpha^2+(1+\alpha^2\beta)^2\Big]^{1/2}\,,
\end{equation}
where $\kappa^2=8\pi G$ and
\begin{equation}\label{AB}
\alpha = {\kappa^2\,S_1\,H_T\over H_F^2-H_T^2}\,, \hspace{5mm}
\beta = {H_F^2-H_T^2\over4H_T^2} \,.
\end{equation}
Here $S_1$ is the contribution to the bounce action coming from the
bubble wall. At tunneling, $\phi=\phi_*$, the potential (\ref{tunnel})
becomes,
\begin{equation}\label{US}
U(\sigma) = {\lambda'\over4}\,\sigma^2 (\sigma-\sigma_0)^2 \,,
\end{equation}
where $\sigma_0 = 2M'/\sqrt{\lambda'}$. Thus $S_1$ can be computed
as~\cite{bubble}
\begin{equation}\label{S1}
S_1 = \int_0^{\sigma_0} d\sigma\,[2(U(\sigma)-U_F)]^{1/2} =
{2\sqrt2 M'^3\over3\lambda'}\,.
\end{equation}

Note that in the limit $\alpha^2\beta\ll1$ of small gravitational effects
at tunneling, we recover the result of Ref.~\cite{LM},
\begin{equation}\label{wall1}
\langle{\cal R}^2_W\rangle =
{3\,H_T^2(H_F^2-H_T^2)\over4\pi^2\kappa^2S_1^2}\,.
\end{equation}
However, in the opposite limit of strong gravitational effects,
$\alpha^2\beta\gg1$, we have~\cite{bubble}
\begin{equation}\label{wall2}
\langle{\cal R}^2_W\rangle = {3\kappa^2\,H_T^2\over16\pi^2}\,.
\end{equation}

\begin{figure}[t]
\centering
\leavevmode\epsfysize=5.78cm \epsfbox{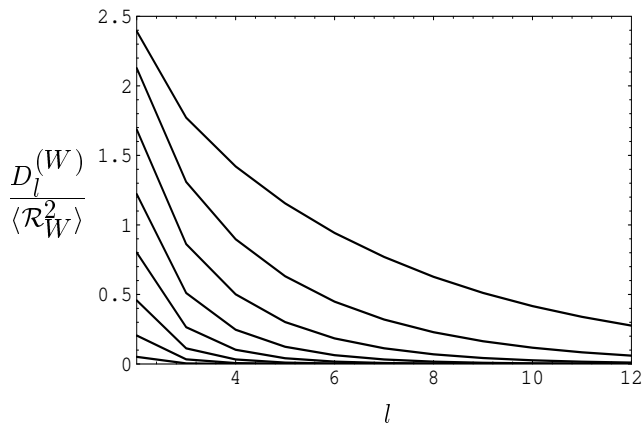}\\
\caption[fig5]{\label{fig5} The first 12 multipoles of the angular power
  spectrum associated with the bubble wall modes, $D_l^{(W)}$,
  normalized to the curvature perturbation amplitude, $\langle{\cal
  R}^2_W\rangle$, for $\Omega_0 = 0.2$ to $0.9$ from top to bottom.
  The quadrupole dominates the $C_l^{(W)}$ power spectrum for all
  $\Omega_0$.}
\end{figure}

The contribution of the bubble wall mode to the angular power spectrum
can be written as
\begin{equation}\label{powerW}
D_l^{(W)} = l(l+1) C_l^{(W)} =
{4\pi\,\langle{\cal R}^2_W\rangle\over(l+2)(l-1)}\,\bar W_{2l}^2\,,
\end{equation}
where $\bar W_{2l}$ is the window function for this $|q|=2$ mode,
see Ref.~\cite{induced} for the correctly normalized eigenfunctions.

We have computed the angular power spectrum $C_l^{(W)}$ for different
values of $\Omega_0$ in the range $0.2\leq\Omega_0\leq0.9$, as shown
in Fig.~5 for the first twelve multipoles. We have found $l(l+1)
C_l^{(W)}/\langle{\cal R}^2_W\rangle$ to be of the same order of
magnitude or smaller than the continuum spectrum, see Fig.~6. Since we
want the bubble wall mode contribution to the power spectrum to be
smaller than that of the continuum modes, we need
\begin{equation}\label{condW}
\langle{\cal R}^2_W\rangle < A^2_C\,.
\end{equation}
This imposes only a very mild constraint on the parameters of the
model. For $\alpha^2\beta\ll1$, we would have
\begin{equation}\label{constrW}
\alpha\,e^{\alpha N/3} < {3g\,S_1\over M(U_F-U_T)^{1/2}}\,,
\end{equation}
which is easy to satisfy for sufficiently large $M'$, see
Ref.~\cite{LM}. However, since we are considering the tunneling when
$H_F \simeq H_T$, we have $\alpha^2\beta\gg1$ and the amplitude of the
metric perturbation (\ref{wall2}) becomes
\begin{equation}\label{Wcon}
\langle{\cal R}^2_W\rangle =
{1\over\lambda}\,{M^4\over M_{\rm P}^4}\,,
\end{equation}
which could indeed be smaller than $A^2_C$ for the parameters of
tilted hybrid models, $\lambda\simeq 0.1$ and $M\simeq 2\times 10^{-3}
M_{\rm P}$, discussed in Section~\ref{DYN}.

\begin{figure}[t]
\centering
\leavevmode\epsfysize=5.64cm \epsfbox{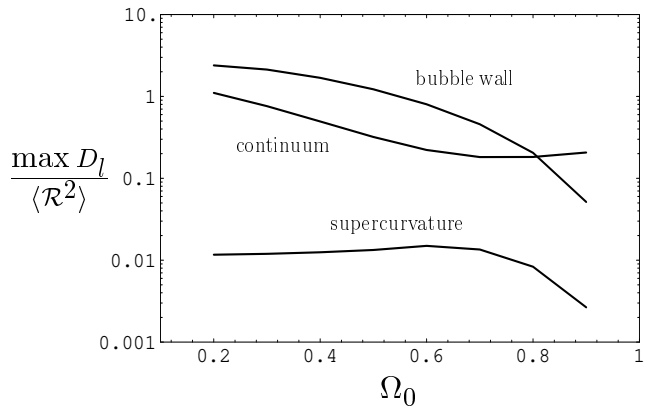}\\
\caption[fig6]{\label{fig6} The largest multipole of the angular power
  spectrum, normalized to the corresponding metric perturbation, for
  the continuum, supercurvature and bubble wall modes, as a function
  of $\Omega_0$. While the continuum modes do not change much as we
  approach $\Omega_0=1$, the supercurvature and bubble wall
  modes decrease exponentially. }
\end{figure}

\section{Conclusions}

In this paper we have studied an open inflation model in which the
second period of expansion inside the bubble is driven by tilted
hybrid inflation~\cite{GBL}. This model has a characteristic
signature of a tilted $n>1$ spectrum. Such a spectrum is nowadays the
best candidate for explaining the CMB anisotropies in case the
universe is wide open, with $\Omega_0 < 0.4$~\cite{WS}.

We have computed the angular power spectrum of temperature
fluctuations in the CMB induced by the continuum of scalar
subcurvature modes, as well as the de Sitter supercurvature mode and
the bubble wall mode. We have found a set of constraints that the
parameters of such a model should satisfy in order to agree with
observations. The parameters of our model satisfy all the constraints
in a natural way. The contribution of a primordial spectrum of
gravitational waves to the complete angular power spectrum of
temperature anisotropies and the associated constraints on the
parameters of the open hybrid model will be discussed in a separate
publication.

Note that our model requires the existence of three scalar fields,
which is a long shot from the simplest inflationary models describing
only one scalar field. On the other hand, our model represents a
simple combination of the hybrid inflation scenario and the theory of
a scalar field $\sigma$ with a metastable vacuum state $\sigma=0$.
The hybrid inflation scenario recently became very popular. It allows
inflation to occur at sub-Planckian values of scalar fields, which
makes it easier to realize it in the context of supersymmetric models.
The scalar field $\sigma$ with a metastable vacuum state is necessary
for open inflation anyway. Thus our model consists of only two
building blocks, each of which is rather natural in the context of
inflationary cosmology.  There are many ways to combine these two
models together, so the model described in this paper is not unique,
and it is quite possible that a simpler and more elegant model can be
proposed. Our main purpose here was simply to show that open inflation
models with blue spectra do exist. On the other hand, the tilted
hybrid inflation which we used in our scenario has a naturally short
duration of inflation after the tunneling with the open universe
formation. This is exactly what we need in order to have an
inflationary universe with $\Omega<1$.

In principle, one can obtain a tilted spectrum of perturbations in an
entirely different way, by combining the open inflation model with the
recently proposed model of Ref.~\cite{LinMukh}, in which one may have
strongly tilted isocurvature or adiabatic non-Gaussian perturbations.
It might be possible to obtain a model with all desirable properties
that will require only two scalar fields.

To avoid misunderstandings, we should emphasize that we sincerely hope
that at the end of the day observational data will show that the
universe is flat, and that perturbations of the metric are
scale-invariant, as suggested by the simplest inflationary models.
However, when the universe was built we have not been consulted. Our
final goal is to propose an internally consistent theory that will
correctly describe observational data. We are very encouraged that a
consistent theory of an open universe does exist. So far this theory
has been formulated only in the context of inflationary cosmology.
This means that the consistent cosmological models describing large
nearly homogeneous universe with $\Omega < 1$ do not contradict
inflation, and, moreover, they have so far been constructed only
within the context of inflation. In this paper we have shown that it
is not difficult to modify the spectrum of perturbations of the metric
in such models, which may be necessary to make them consistent with
observations.

\section*{Acknowledgements}
The work of A.L.  was supported by NSF grant PHY-9219345. This work
was also supported by a NATO Collaborative Research Grant,
Ref.~CRG.950760.


\begin{references}

\bibitem{open} J. R. Gott, Nature, {\bf 295}, 304 (1982); J. R. Gott III and  
T. S. Statler, Phys Lett {\bf 136B}, 157 (1984).
 
 \bibitem{BGT} M.  Bucher, A. S. Goldhaber and N. Turok, Phys. Rev. D
{\bf 52}, 3314 (1995); M.  Bucher and N. Turok, Phys. Rev. D
{\bf 52}, 5538 (1995); K. Yamamoto, M. Sasaki and T. Tanaka,
Astrophys. J. {\bf 455}, 412 (1995).


 \bibitem{LM} A. D. Linde, Phys. Lett. {\bf B 351}, 99 (1995); A. D.
Linde and A. Mezhlumian, Phys. Rev. D {\bf 52}, 6789 (1995).

\bibitem{HST} W. L. Freedman et al., Nature {\bf 371}, 757
  (1994); N. R. Tanvir et al., Nature {\bf 377}, 27 (1995).

\bibitem{GC} M. Bolte and C. J. Hogan, Nature {\bf 376}, 399 (1995).
  
\bibitem{LVIL} A.D. Linde, JETP {\bf 60}, 211 (1984); Lett. Nuovo Cim.
  {\bf 39}, 401 (1984); Ya.B. Zeldovich and A.A. Starobinsky, Sov.
  Astron.  Lett. {\bf 10}, 135 (1984); V.A. Rubakov, Phys. Lett. {\bf
    148B}, 280 (1984); A. Vilenkin, Phys. Rev. {\bf D30}, 549 (1984).

\bibitem{sasaki} M. Sasaki, T. Tanaka and K. Yamamoto, Phys. Rev.  D
  {\bf 51}, 2979 (1995).

\bibitem{LW} D. H. Lyth and A. Woszczyna, Phys. Rev. D {\bf 52}, 3338
  (1995).

\bibitem{GZ} J. Garc\'{\i}a-Bellido, A. R. Liddle, D. H. Lyth, and D.
  Wands, Phys. Rev. D {\bf 52}, 6750 (1995).

\bibitem{YST} K. Yamamoto, M.  Sasaki and T. Tanaka, Phys. Rev. D {\bf
    54}, 5031 (1996).

\bibitem{Hamazaki} T. Hamazaki, M. Sasaki, T. Tanaka and K. Yamamoto,
  Phys. Rev. D {\bf 53}, 2045 (1996).

\bibitem{bubble} J. Garc\'{\i}a-Bellido, Phys. Rev. D {\bf 54}, 2473
  (1996).

\bibitem{Garriga} J. Garriga, Phys. Rev. D {\bf 54}, 4764, (1996).

\bibitem{super} M. Sasaki and T. Tanaka, Phys. Rev. D {\bf 54}, R4705
  (1996).

\bibitem{WS} M. White and J. Silk, Phys. Rev. Lett. {\bf 77}, 4704
  (1996), {\tt astro-ph/9608177}.

\bibitem{induced} J. Garc\'{\i}a-Bellido and A. R. Liddle, {\it Complete
    Power Spectrum for an Induced Gravity Open Inflation Model}, {\tt
    astro-ph/9610183} (1996).

\bibitem{hybrid} A.D. Linde, Phys. Lett. {\bf B259}, 38 (1991); Phys.
  Rev. D {\bf 49}, 748 (1994).

\bibitem{LL93} A. R. Liddle and D. H. Lyth, Phys. Rep. {\bf 231}, 1
  (1993).

\bibitem{CLLSW} E. J. Copeland, A. R. Liddle, D. H. Lyth,
  E. D.  Stewart and D. Wands, Phys. Rev. D {\bf 49}, 6410 (1994).


\bibitem{GBW} J. Garc\'\i a-Bellido and D. Wands, Phys.
  Rev. D {\bf 54}, 7181, (1996).

\bibitem{Lyth} For a comprehensive review, see D.H. Lyth, {\it Models of
  Inflation and the Spectral Index of the Density Perturbation,} {\tt
    hep-ph/9609431} (1996).

\bibitem{GBL} J. Garc\'\i a-Bellido and A. Linde, {\it Tilted
  Hybrid Inflation}, {\tt astro-ph/9612141} (1996).

\bibitem{TS} T. Tanaka and M. Sasaki, Osaka preprint, OU-TAP 48 {\tt
    astro-ph/9701053} (1997); M. Bucher and J. D. Cohn, {\it Primordial
  gravitational waves from open inflation}, {\tt astro-ph/9701117}
  (1996).

\bibitem{Guth} L. Randall, M. Solja\v ci\'c and A. H. Guth, Nucl.
  Phys. {\bf B 472}, 377 (1996).

\bibitem{GBLW} J. Garc\'\i a-Bellido, A. D. Linde and D. Wands, Phys.
  Rev. D {\bf 54}, 6040 (1996).

\bibitem{LinMukh} A. Linde and V. Mukhanov, {\it NonGaussian
    Isocurvature Perturbations from Inflation}, {\tt astro-ph/9610219}
  (1996).

\bibitem{moduli} M. Dine, W. Fischler and D. Nemeschansky, Phys.
  Lett. {\bf B136}, 169 (1984); G.D. Coughlan, R. Holman, P.  Ramond
  and G.G. Ross, Phys. Lett. {\bf B140}, 44 (1984); A.S. Goncharov,
  A.D.  Linde and M.I. Vysotsky, Phys. Lett.  {\bf 147B}, 279 (1984);
  M. Dine, L.  Randall and S. Thomas, Nucl.  Phys. {\bf 458}, 291
  (1996).

\bibitem{MAP} {\it MAP} Home Page at \\
  {\tt http://map.gsfc.nasa.gov/} (1996).

\bibitem{COBRAS} {\it COBRAS/SAMBA} Home Page at \\
   {\tt http://astro.estec.esa.nl/SA-general/Projects/\\
   Cobras/cobras.html} (1996).

\bibitem{Cohn} J. D. Cohn, Phys. Rev. D {\bf 54}, 7215, (1996).

\bibitem{SW67} R. K. Sachs and A. M. Wolfe, Astrophys. J. {\bf 147},
  73 (1967).

\bibitem{Har} E. R. Harrison, Rev. Mod. Phys. {\bf 39}, 862 (1967); M.
  L. Wilson, Astrophys. J.  {\bf 273}, 2 (1983); L. F. Abbott and R.
  K. Schaefer, Astrophys. J.  {\bf 308}, 546 (1986).

\bibitem{peak} W. Hu and N. Sugiyama, Phys. Rev. D {\bf 51}, 2599 (1995).

\bibitem{future} G. Jungman, M. Kamionkowski, A. Kosowsky and
  D. N. Spergel, Phys. Rev. D {\bf 54}, 1332 (1996).

\bibitem{COBE} C. L. Bennett et al., Astrophys. J. {\bf 464}, L1 (1996).

\bibitem{BW}   E.F. Bunn, A.R. Liddle and  M. White,  Phys. Rev. D {\bf 54},
5917 (1996), {\tt astro-ph/9607038}.

\bibitem{modes} J. Garc\'{\i}a-Bellido, A. R. Liddle, D. H. Lyth and
  D. Wands, {\it Normalization of Modes in an Open Universe}, {\tt
    astro-ph/9608106} (1995).

\end{references}
\end{document}